\documentclass[aps,prb,twocolumn,showpacs,superscriptaddress]{revtex4}

\usepackage{graphicx} 

\begin{document}

\title
{A systematic ab-initio study of curvature effects
in carbon nanotubes}

\author{O. G\"{u}lseren}
\affiliation{NIST Center for Neutron Research,
National Institute of Standards and Technology,
Gaithersburg, MD 20899}
\affiliation{Department of Materials Science and Engineering,
University of Pennsylvania, Philadelphia, PA 19104}
\author{T. Yildirim}
\affiliation{NIST Center for Neutron Research,
National Institute of Standards and Technology,
Gaithersburg, MD 20899}
\author{S. Ciraci}
\affiliation{Department of Physics, Bilkent University,
Ankara 06533, Turkey}

\date{\today}

\begin{abstract}
We investigate curvature effects on geometric parameters, energetics
and electronic structure of zigzag nanotubes with fully optimized
geometries from first-principle calculations. The calculated curvature
energies, which are inversely proportional to the square of radius,
are in good agreement with the classical elasticity theory.
The variation of the band gap with radius is found to differ from
simple rules based on the zone folded graphene bands. Large
discrepancies between tight binding and first principles calculations
of the band gap values of small nanotubes are discussed in detail.
\end{abstract}



\maketitle

\section{Introduction}

Single wall carbon nanotubes (SWNTs) are basically rolled graphite
sheets, which are characterized by two integers $(n,m)$ defining the
rolling vector of graphite.\cite{dressel} Therefore, electronic
properties of SWNTs, at first order, can be deduced from that of
graphene by mapping the band structure of 2D hexagonal lattice on
a cylinder.\cite{dressel,hamada,dressel1,mintmire2,mintmire3}
Such analysis indicates that the $(n,n)$ armchair nanotubes are always
metal and exhibit one dimensional quantum conduction~\cite{heer}.
The $(n,0)$ zigzag nanotubes are generally semiconductor and only are
metal if $n$ is an integer multiple of three. However, recent
experiments\cite{lieber} indicate much more complicated structural
dependence of the band gap and electronic properties of SWNTs.
The semiconducting behavior of SWNTs has been of particular interest,
since the electronic properties can be controlled by doping or
implementing defects in a nanotube-based optoelectronic
devices.\cite{chico,collins,bockra,bezryadin,cetin,hssim,tubeadb}
It is therefore desirable to have a good understanding of electronic
and structural properties of SWNTs and the interrelations between them.

Band calculations of SWNTs were initially performed by using a
one\textendash band $\pi$\textendash orbital tight binding
model.\cite{hamada} Subsequently, experimental
data~\cite{dekker,venema,odom1,odom2} on the band gaps were extrapolated
to confirm the inverse proportionality with the radius of the
nanotube.\cite{mintmire3} Later, first principles calculation~\cite{blase}
within Local Density Approximation (LDA) showed that the
$\sigma^*$\textendash $\pi^*$ hybridization becomes significant at small
$R$ (or at high curvature). Such an effect were not revealed by the
$\pi$\textendash orbital tight-binding bands. Recent analytical
studies~\cite{mele,klein1,klein2} showed the importance of curvature
effects in carbon nanotubes.  Nonetheless, band calculations performed
by using different methods have been at variance on the values of the
band gap. While recent studies predict interesting effects, such as
strongly local curvature dependent chemical reactivity~\cite{tubeadb},
an extensive theoretical analysis of the curvature effects on geometric
and electronic structure has not been carried out so far.

In this paper, we present a systematic ab-initio analysis of the band 
structure of zigzag SWNTs showing interesting curvature effects. Our
analysis includes a large number of zigzag SWNTs with $n$ ranging from
4 to 15. The fully optimized structural and electronic properties
of SWNTs are obtained from extensive first-principle calculations
within the Generalized Gradient Approximation\cite{gga} (GGA) by using
pseudopotential planewave method~\cite{castep}. We used plane waves
up to an energy of 500 eV and ultrasoft pseudopotentials~\cite{usps}.
The calculated  total energies converged within 0.5 meV/atom.
More details about the calculations can be found in
Ref's[\onlinecite{taner,oguz2}].

\section{Geometric Structure}
\label{sec:idealgeom}

\begin{figure}
\includegraphics[scale=0.38]{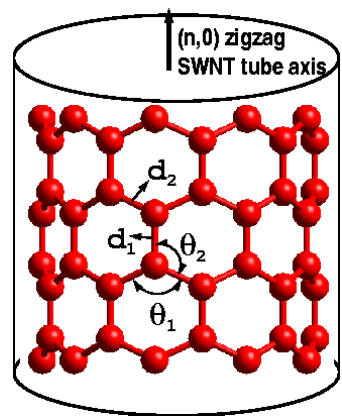}
\caption{A schematic side view of a zigzag SWNT, indicating two types
of C\textendash C bonds and C\textendash C\textendash C bond angles.
These are labeled as $d_{1}$, $d_{2}$, $\theta_{1}$ and $\theta_{2}$.
Radius dependence of these variables are important in tight-binding
description of SWNTs as discussed in the text.}
\label{fig:zigzagstr}
\end{figure}

\begin{figure}
\includegraphics[scale=0.5]{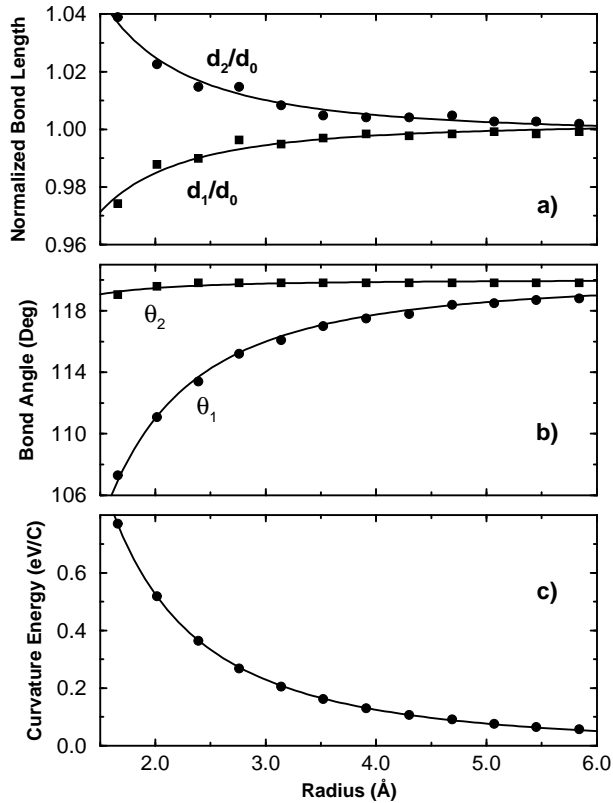}
\caption{(a) Normalized bond lengths ($d_{1}/d_0$ and $d_{2}/d_0$)
versus the tube radius $R$. ($d_0=1.41$~\AA{}).
(b) The bond angles ($\theta_{1}$ and $\theta_{2}$) versus $R$.
(c) The curvature energy, $E_c$ per carbon atom with respect to graphene
as a function of tube radius.
The solid lines are the fit to the data as $1/R^2$.}
\label{fig:idstr}
\end{figure}

First, we discuss effects of curvature on structural parameters such
as bond lengths and angles. Figure~\ref{fig:zigzagstr} shows a schematic
side view of a zigzag SWNT which indicates two types of C\textendash C
bonds and C\textendash C\textendash C bond angles, respectively.
The curvature dependence of the fully optimized structural parameters
of zigzag SWNTs are summarized in Fig.~\ref{fig:idstr}. The variation of
the normalized bond lengths (\textit{i.e.} $d_{C-C}/d_{0}$ where $d_0$
is the optimized C\textendash C bond length in graphene) and the bond
angles with tube radius $R$ (or $n$) are shown in
Fig.~\ref{fig:idstr}a and b, respectively. Both the bond lengths and
the bond angles display a monotonic variation and approach the graphene
values as the radius increases. As pointed out earlier for the armchair
SWNTs~\cite{portal}, the curvature effects, however, become significant
at small radii. The zigzag bond angle ($\theta_{1}$) decreases with
decreasing radius. It is about  $12^o$ less than $120^o$,  namely the
bond angle between $sp^2$ bonds of the graphene,  for the $(4,0)$ SWNT,
the smallest tube we studied. The length of the corresponding zigzag
bonds ($d_{2}$), on the other hand, increases with decreasing $R$.
On the other hand, the length of the parallel bond ($d_{1}$) decreases
to a lesser extent with decreasing $R$, and the angle involving this
bond ($\theta_{2}$) is almost constant.

An internal strain is implemented upon the formation of tubular structure
from the graphene sheet. The associated strain energy, which is specified
as the curvature energy, $E_c$, is calculated as the difference of total
energy per carbon atom between the bare SWNT and the graphene
(\textit{i.e.} $E_c=E_{T,SWNT} - E_{T,graphene}$) for $4\leq n \leq 15$.
The calculated curvature energies are shown in Fig.~\ref{fig:idstr}c.
As expected $E_{c}$ is positive and increases with increasing curvature.
Consequently, the binding (or cohesive) energy of carbon atom in a SWNT
decrease with increasing curvature. We note that in the classical theory
of elasticity the curvature energy is given by the following
expression~\cite{mintmire1,tibbets,kudin}
\begin{equation}
E_{c}=\frac{Y h^3}{24}\frac{\Omega}{R^2}=\frac{\alpha}{R^2} .
\label{idest}
\end{equation}
Here $Y$ is the Young's modulus, $h$ is the thickness of the tube,
and $\Omega$ is the atomic volume. Interestingly, the ab-initio curvature
energies yield a perfect fit to the relation $\alpha/R^2$ as seen in
Fig.~\ref{fig:idstr}c. This situation suggests that the classical theory
of elasticity can be used to deduce the elastic properties of SWNTs.
In this fit $\alpha$ is found to be 2.14~eV \AA$^2$/atom, wherefrom $Y$
can be calculated with an appropriate choice of $h$.

\section{Electronic Structure}
\label{sec:idealelec}

\begin{table*}
\caption{Band gap, $E_g$, as a function of radius $R$ of (n,0) zigzag
nanotubes. M denotes the metallic state. Present results for $E_g$ were
obtained within GGA. First row of Ref.~\protect\onlinecite{blase} is LDA
results while all the rest are tight-binding (TB) results. Two rows of
Ref.~\protect\onlinecite{yori2} are for two different TB parametrization.}
\label{table:ideg}
\begin{ruledtabular}
\begin{tabular}{l|cccccccccccc}
$n$ & 4 & 5 & 6 & 7 & 8 & 9 & 10 & 11 & 12 & 13 & 14 & 15 \\
\hline
R (\AA) & 1.66 & 2.02 & 2.39 & 2.76 & 3.14 & 3.52 & 3.91 & 4.30 & 4.69 &
5.07 & 5.45 & 5.84 \\
\hline
$E_g$ (eV) & M & M & M & 0.243 & 0.643 & 0.093 & 0.764 & 
0.939 & 0.078 & 0.625 & 0.736 & 0.028 \\
Ref.~\onlinecite{blase}  &  &  & M & 0.09 & 0.62  & 0.17  &  &  &  &  &  &  \\
Ref.~\onlinecite{blase} &  &  & 0.05  & 1.04 & 1.19 & 0.07 &  &  &  &  &  &  \\
Ref.~\onlinecite{hamada} &  &  & 0.21  & 1.0 & 1.22 & 0.045 & 0.86 & 0.89 &
 0.008 & 0.697 & 0.7 & 0.0 \\
Ref.~\onlinecite{yori2} &  &  &   & 0.79 & 1.12 &  & 0.65 & 0.80 &  &  &  &  \\
Ref.~\onlinecite{yori2} &  &  &   & 1.11 & 1.33 &  & 0.87 & 0.96 &  &  &  &  \\
\end{tabular}
\end{ruledtabular}
\end{table*}

An overall behavior of the electronic band structures of SWNTs has been
revealed from zone folding of the graphene
bands.\cite{hamada,dressel1,mintmire2}
Accordingly, all $(n,0)$ zigzag SWNT were predicted to be metallic when
$n$ is multiple of 3, since the double degenerate $\pi$ and $\pi^*$
states, which overlap at the $K$\textendash point of the hexagonal
Brillouin zone (BZ) of graphene folds to the $\Gamma$ point of the
tube~\cite{hamada,mintmire2}. This simple picture provides a qualitative
understanding, but fails to describe some important features, in
particular for small radius or \textit{metallic} nanotubes. This is
clearly shown in  Table~\ref{table:ideg}, where the band gaps calculated
in the present study are summarized and compared with results obtained
from other methods in the literature. For example, our calculations result
in small but non-zero energy band gaps of 93, 78 and 28 meV for $(9,0)$,
$(12,0)$ and $(15,0)$ SWNTs, respectively (see Table~1). Recently,
these gaps are measured by Scanning Tunneling Spectroscopy (STS)
experiments~\cite{lieber} as 80, 42 and 29 meV, in the same order.
The biggest discrepancy noted in Table~\ref{table:ideg} is between the
tight-binding and the first-principles values of the gaps for small radius
tubes such as $(7,0)$. These results indicate that curvature effects are
important and the simple zone folding picture has to be improved.
Moreover, the analysis of the LDA bands of the $(6,0)$ SWNT calculated
by Blase \textit{et al}~\cite{blase} brought another important effect
of the curvature. The antibonding singlet $\pi^*$ and $\sigma^*$ states
mix and repel each other in curved graphene. As a result, the purely
$\pi^*$ state of planar graphene is lowered with increasing curvature.
For zigzag SWNTs, the energy of this singlet $\pi^*$\textendash state is
shifted downwards with decreasing $R$ (or increasing curvature). Here,
we extended the analysis of Blase \textit{et al}~\cite{blase} to the
$(n,0)$ SWNTs with $4\leq n \leq 15$ by performing GGA calculations.

\begin{figure}
\includegraphics[scale=0.5]{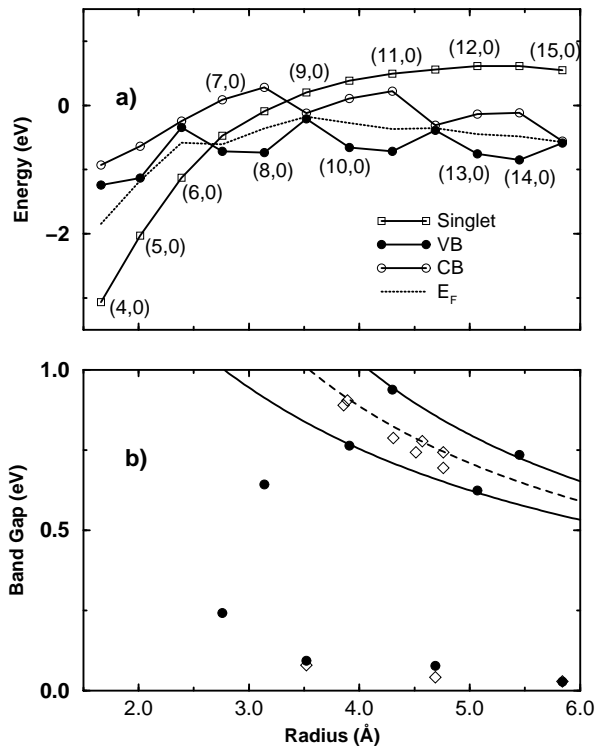}
\caption{(a) Energies of the double degenerate $\pi$\textendash states
(VB), the double degenerate $\pi^*$\textendash states (CB) and the
singlet $\pi^*$\textendash state as a function of nanotube radius.
Each data point corresponds to $n$ ranging from 4 to 15 consecutively.
(b) The calculated band gaps as a function of the tube radius
shown by filled symbols. Solid (dashed) lines are the plots of
Eq.~\protect\ref{gap1r2} (Eq.~\protect\ref{gap1r}). The experimental
data are shown by open diamonds~\protect\cite{odom1,odom2,lieber}.} 
\label{fig:idband}
\end{figure}

In Fig.~\ref{fig:idband}a, we show the double degenerate
$\pi$\textendash states (which are the valence band edge at the
$\Gamma$\textendash point), the double degenerate $\pi^*$\textendash states
(which become the conduction band edge at $\Gamma$ for large $R$), and the
singlet $\pi^*$\textendash state (which is in the conduction band for large
$R$). As seen, the shift of the singlet $\pi^*$\textendash  state is
curvature dependent, and below a certain radius determines the band gap.
For tubes with radius greater than 3.3 \AA{} (\textit{i.e.} $n > 8$),
the energy of the singlet $\pi^*$\textendash state at the
$\Gamma$\textendash point of the BZ is above the doubly degenerate
$\pi^*$ states (\textit{i.e.} bottom of the conduction band), while it
falls between the valence and conduction band edges for $n=7,8$, and
eventually dips even below the double degenerate valence band
$\pi$\textendash states for the zigzag SWNT with radius less than 2.7 \AA{}
(\textit{i.e.} $n < 7$). Therefore, all the zigzag tubes with radius less
than 2.7 \AA{} are metallic. For $n=7,8$, the edge of the conduction band
is made by the singlet $\pi^*$\textendash state, but not by the double
degenerate $\pi^*$\textendash state. The band gap derived from the zone
folding scheme is reduced by the shift of this singlet
$\pi^*$\textendash state as a result of curvature induced
$\sigma^* - \pi^*$ mixing. This explains why the tight binding
calculations predict band gaps around 1~eV for $n=7,8$ tubes while the
self-consistent calculations predict much smaller value.

Another issue we next address is the variation of the band gap, $E_g$,
as a function of tube radius. Based on $\pi$\textendash orbital tight
binding model, it was proposed~\cite{mintmire3} that $E_g$ behaves as
\begin{equation}
E_{g}=\gamma_0 \frac{d_0}{R},
\label{gap1r}
\end{equation}
which is independent from helicity. Within the simple
$\pi$\textendash orbital tight binding model, $\gamma_{o}$ is taken to be
equal to the hopping matrix element $V_{pp\pi}$. ($d_0$ is the bond
length in graphene). 
However, as seen in Fig.~\ref{fig:idband}b, the band gap displays a rather
oscillatory behavior up to radius 6.0 \AA. The relation given in
Eq.~\ref{gap1r} was obtained by a second order Taylor expansion of
one-electron eigenvalues of $\pi$-orbital tight binding
model~\cite{mintmire3} around the $K$\textendash point of the BZ, and
hence it fails to represent the effect of the helicity. By extending
the Taylor expansion to the next higher order, Yorikawa and
Muramatsu~\cite{yori1,yori2} included another term in the empirical
expression of the band gap variation,
\begin{equation}
E_{g} = V_{pp\pi} \frac{d_0}{R} 
        [ 1 + (-1)^p \gamma cos(3\theta) \frac{d_0}{R} ],
\label{gap1r2}
\end{equation}
which depends on the chiral angle, $\theta$, as well as an index $p$.
Here $\gamma$ is a constant and the index $p$ is defined as the integer
from $k=n-2m=3q+p$. The factor $(-1)^p$ comes from the fact that the
allowed $\mathbf{k}$ is nearest to either the $K$- or $K^{'}$- point of
the hexagonal Brillouin zone. For zigzag nanotubes studied here, the
chiral angle is zero, so the second term just gives $R^{-2}$ dependence
as $\pm \gamma V_{pp\pi} (d_0/R)^2$. Hence, the solid lines in
Fig.~\ref{fig:idband}b are fits to the empirical expression,
$E_{g}=V_{pp\pi} d_0/R \pm V_{pp\pi} \gamma d^2_0/R^2$, obtained from
Eq.~\ref{gap1r2} for $\theta=$0 by using the parameters $V_{pp\pi}=2.53$ eV
and $\gamma=0.43$. The experimental data obtained by STS~\cite{odom1,odom2}
are shown by open diamonds in the same figure. The agreement between our
calculations and the experimental data is very good considering the fact
that there might be some uncertainties in identifying the nanotube
(\textit{i.e.} assignment of $(n,m)$ indices) in the experiment.
The fit of this data to the empirical expression given by Eq.~\ref{gap1r}
are also presented by a dashed line for comparison.

The situation displayed in Fig.~\ref{fig:idband} indicates that
the variation of the band gap with the radius is not simply $1/R$, but
additional terms incorporating the chirality dependence are required.
Most importantly, the mixing of the singlet $\pi^*$\textendash state with
the the singlet $\sigma^*$\textendash state due to the curvature, and its
shift towards the valence band with increasing curvature is not included
in neither the $\pi$\textendash orbital tight binding model, nor the
empirical relations expressed by Eq.~\ref{gap1r} and \ref{gap1r2}.
This behavior of the singlet $\pi^*$\textendash states is of particular
importance for the applied radial deformation that modifies the curvature
and in turn induces metallization~\cite{cetin,oguz2,park}.

In conclusion, we investigated structural and electronic properties that
result from the tubular nature  of the SWNTs. The first-principles total
energy calculations indicated that significant amount of strain energy
is implemented in a SWNT when the radius is small. However, the elastic
properties can be still described by the classical theory of elasticity.
We showed how the singlet $\pi^{*}$\textendash state in the conduction
band of a zigzag tube moves and eventually enters in the band gap
between the doubly degenerate $\pi^{*}$-conduction and $\pi$-valence
bands. As a result, the energy band structure and the variation of the
gap with radius (or $n$) differs from what one derived from the zone
folded band structure of graphene based on the simple tight binding
calculations.

\textbf{Acknowledgments}
This work was partially supported by the NSF under Grant No. INT01-15021
and T\"{U}B\'{I}TAK under Grant No. TBAG-U/13(101T010).

\end{document}